 \documentclass[final,5p,times,twocolumn]{elsarticle}

\usepackage{graphicx,stfloats}
\usepackage{color}

\usepackage{graphicx}
\usepackage{longtable}
\usepackage{tabularx}
\usepackage{url}
\usepackage{subcaption}
\usepackage{lipsum} 
\usepackage{amssymb,amsmath}
\usepackage[version=3]{mhchem}

\newcounter{notecount}

\graphicspath{{../figs/}}

\def\one{\columnwidth}
\def\two{1.95\columnwidth}

\biboptions{sort&compress}

\journal{Combustion and Flame}

\begin{document}

\begin{frontmatter}



\title{Efficient and accurate calculation of dispersion relations for intrinsically unstable premixed flames}


\author[uoe]{Sofiane Al Kassar\corref{cor1}}
\cortext[cor1]{Corresponding author}
\ead{S.Al-Kassar@sms.ed.ac.uk}
\author[rwth]{Lukas Berger}
\author[rome]{Pasquale E. Lapenna}
\author[rome]{Francesco Creta}
\author[rwth]{Heinz Pitsch}
\author[uoe]{Antonio Attili}

\affiliation[uoe]{organization={School of Engineering, Institute for Multiscale Thermofluids, University of Edinburgh},
            addressline={}, 
            city={Edinburgh},
            postcode={EH9 3FD}, 
            state={},
            country={United Kingdom}}

\affiliation[rwth]{organization={Institute for Combustion Technology, RWTH Aachen University},
            addressline={}, 
            city={Aachen},
            postcode={52062}, 
            state={},
            country={Germany}}

 \affiliation[rome]{organization={DIMA, Sapienza, University of Rome},
            addressline={}, 
            city={Rome},
            postcode={00189}, 
            state={},
            country={Italy}}

\begin{abstract}
Premixed flames are susceptible to hydrodynamic and thermodiffusive instabilities that wrinkle the 
flame front and lead to complex multiscale patterns. They
strongly impact the flame propagation and dynamics, increasing the speed of a laminar flame 
by several folds, easily as large as a factor of five for lean hydrogen flames at high pressure.  
The dispersion relation, which represents the growth rate of the different harmonic components of the perturbation of the flame front for different wavelengths, is useful
to understand the dynamics during the linear phase of flame instabilities.
In this work, an efficient and accurate approach based on a Fourier analysis of flame wrinkling 
is proposed to calculate the dispersion relation.
Differently from the typical approach based on perturbing the flame with a single wavelength, the flame is perturbed with a spectrum of 
sine waves and their growth is followed with a spectral analysis. 
With the present method, the full dispersion relation is computed with a single simulation; this is 
significantly more efficient computationally than running a series of simulations with a single-wavelength 
perturbation for each point of the dispersion relation. It is shown that the presented approach is accurate and also solves an issue
encountered when a single perturbation is imposed to compute the growth rate of large wavelengths.
Several numerical and initialisation parameters, including resolution, domain size, and amplitude of the initial 
perturbation, are studied systematically and assessed.

\end{abstract}




\end{frontmatter}


\section*{Novelty and Significance Statement}

This paper presents a new approach to compute the entire dispersion relation of premixed flames with a single simulation, whereas previous work required several simulations. It provides clear guidelines for selecting simulation parameters, including resolution, initial perturbation amplitude, and domain size. It is shown that the method is accurate and robust. The method also proved effective for calculating the dispersion relation for low wavenumbers, which is often difficult.

\section*{Author Contributions}

S.A.K. performed research and wrote the paper, L.B. performed research, P.E.L. performed research, F.C. performed research, H.P. performed research, and A.A. designed and performed research, and wrote the paper.

\section{Introduction}
\label{sec:intro}

Premixed flames subject to perturbations are prone to different intrinsic instabilities~\citep{matalon2007intrinsic,CRETA2020256},
caused by thermal expansion and, potentially, differential transport of mass and heat. 
In the hydrodynamic, or Darrieus-Landau (DL), instability~\citep{darrieus1938propagation,landau1944theory},
the flow acceleration caused by heat release coupled with local flame curvature induces non-uniform flow velocity along the flame front,
causing the growth of the front perturbations. 
Thermodiffusive instabilities are caused by the strong differential diffusion in the combustion of fuels with non-unity Lewis numbers. The discrepancy between the fuel mass and heat fluxes induces a non-uniform distribution of fuel concentration and local flame speed along the flame, which amplifies the front perturbations~\citep{berger2019characteristic}.



Flame instabilities are characterised by two distinct phases. The initial phase~\citep{BERGER2022111935} is characterised by a linear evolution of  
the initial small harmonic perturbation of the flame front, which grows in a purely exponential manner. When the perturbation reaches an 
amplitude of the order of the flame thickness, the instability transition to a non-linear phase~\citep{BERGER2022111936}, 
characterised by complex, multiscale, and often chaotic structures and dynamics. 

The characteristics of the linear phase are usually expressed with a dispersion relation.
It summarises the growth rate $\omega(k)$ of perturbations over a range of wavelengths $\lambda$ or wavenumbers $k=2\pi/\lambda$. The dispersion relation for premixed hydrocarbon flames usually has an approximately parabolic shape, while the shape deviates significantly from this for hydrogen 
flames~\citep{matalon_cui_bechtold_2003, sivashinsky1977nonlinear, BERGER2022111935}.
The different features of the dispersion relation provide important information to understand instabilities, both 
for theoretical analysis~\citep{matalon2007intrinsic,sivashinsky1977nonlinear} and for the development of simplified 
models~\citep{HOWARTH2022111805,berger2019characteristic,BERGER2022111936}. Therefore, efficient and accurate computation of dispersion 
relations is essential, especially considering the recent interest in hydrogen combustion, where thermodiffusive instabilities
play a significant role~\citep{berger2019characteristic,BERGER2022112254}.

A straightforward approach to construct the dispersion relation is to initialise a two-dimensional planar premixed flame and perturb it with a sine wave with a small amplitude and wavenumber $k$. To optimise the simulation, the size of the domain in the spanwise direction is usually set equal to $\lambda=2\pi/k$. The evolution of the sine wave is monitored, and the growth of its amplitude is computed. The dispersion relation $\omega(k)$ is computed by repeating the simulation for different values of $k$~\citep{YUAN20071267, altantzis_2012, FROUZAKIS20151087, KADOWAKI2005193, sharpe_2003, denet_1995, ATTILI20211973, BERGER2022111935, LAPENNA20191945}. This procedure can be rather tedious and computationally expensive since many simulations are required, particularly if a high resolution in wavenumber space is required.




The approach proposed in the paper aims to drastically reduce the computational cost of calculating the dispersion relation. In particular, the goal is to compute the whole dispersion relation with a single simulation.
In addition, we aim to resolve an issue related to the computation of the dispersion relation for small values of the wavenumber $k$, i.e., values of $k$ that are significantly smaller than the wavenumber of peak growth rate. 
In this case, the method based on a single wavelength perturbation is not straightforward. Smaller wavelength (larger wavenumber) components emerge and grow from numerical noise if an accurate and low-dissipative code is used. These components quickly grow larger than the explicitly initialised one, compromising its clear identification.

The paper is organised as follows: Sec.~\ref{sec:Method} describes the proposed method and its application to a lean hydrogen flame. In Sec.~\ref{sec:comparison_single_multi}, the results obtained with the proposed multi-wavelength method are compared with the single-wavelength approach, while the effects of the numerical and initialisation parameters are assessed in Sec.~\ref{sec:param}. A short mathematical analysis of the the method efficiency is given Sec.~\ref{sec:eff} and conclusions are presented in Sec.~\ref{sec:concl}.

\section{Method}
\label{sec:Method}

The approach is based on the spectral analysis of a planar, two-dimensional, premixed flame, initially perturbed with a set of harmonic oscillations with specified wavelengths and amplitudes. A similar approach, based on multi-wavelength excitation impulses, has been used to compute flame transfer functions with a single simulation~\citep{SILVA20153370}. The spectral analysis is based on the Fourier transform of isolines of temperature, or the mass fraction of one of the chemical species, at different time instants. This is used to compute the growth rate of each harmonic component and evaluate the full dispersion relation with a single simulation run. 
This approach is possible since the first phase of the evolution of an unstable flame is linear if the initial perturbations are small enough; under these conditions, the evolution of each wavelength does not depend on the others, and the growth rate of each of them can be computed independently. 
The initial amplitude of the perturbation, the spectrum of wavelengths considered, and the numerical parameters, such as the domain size and grid resolution, are carefully assessed to develop an efficient, accurate, and reliable method for the computation of dispersion relations.
In the rest of the paper, this approach will be referred to as multi-wavelength perturbation method, in contrast with the more traditional method based on a single-wavelength perturbation.

\subsection{Physical models and numerical methods}
The stability properties and the dynamics of a hydrogen flame, characterised by a strong thermodiffusive instability, are investigated by Direct Numerical Simulation (DNS) employing finite rate chemistry and a multistep mechanism to describe chemical reactions. 
The reactive, unsteady Navier-Stokes equations are solved in the low Mach number limit, and the mixture obeys the ideal gas equation of state.
The equation for the full set of chemical species and 
temperature~\citep{berger2019characteristic,attili2016effects,Attili20_ar,berger2022synergistic} are solved employing a finite-rate multistep model with 9 species~\citep{burke2012comprehensive}. The transport properties are
evaluated using a mixture averaged model for the temperature equation~\citep{attili2016effects} while the species diffusivities are computed prescribing spatially homogeneous values of the species Lewis numbers $Le_i$.
The equations are solved with a semi-implicit finite difference code~\citep{desjardins_high_2008}, which has been used and validated in many different configurations~\citep{attili2014formation,attili2016effects,berger2019characteristic,niemietz2022direct}.
Spatial derivatives are discretised with second-order finite differences for the momentum equation and the scalar diffusive terms, while a third-order WENO scheme~\citep{liu1994weighted} is used for the convective term in the scalar equations. The Strang operator splitting is applied to the chemical source term, which is integrated with  the stiff ODE solver CVODE~\citep{hindmarsh2005sundials}.

\subsection{Configuration and flame conditions}

\begin{figure}[h]
    \centering
    \includegraphics[width=\one]{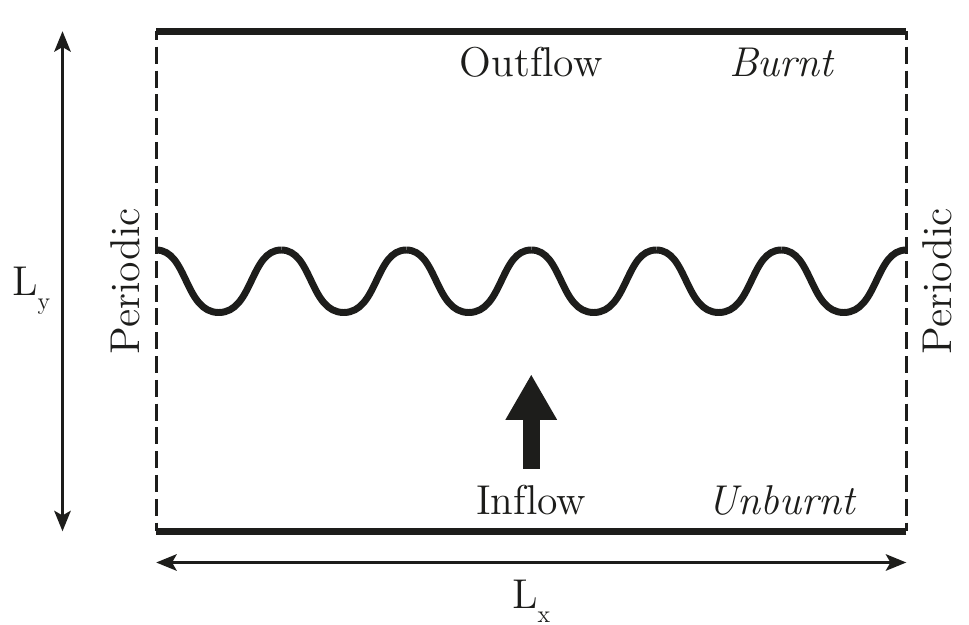}
    \caption{Diagram of the simulation domain and boundary conditions.} 
    \label{fig:diagram_domain}
\end{figure}

A fully-developed planar laminar flame is initialised in a two-dimensional domain sufficiently large in the streamwise direction to allow an unconfined flame evolution. The flow is periodic in the spanwise direction $x$, while inflow and outflow boundary conditions are specified at the two boundaries in the streamwise direction $y$, as represented in Fig.~\ref{fig:diagram_domain}. Temperature and species mass fractions, computed with the FlameMaster code~\citep{pitsch1998flamemaster} in a one-dimensional laminar flame, are mapped in the two-dimensional domain to initialise the simulation. 

To assess the proposed approach, a hydrogen flame at an equivalence ratio of $\phi=0.5$, atmospheric pressure $p=1$~atm, and temperature of the unburnt gas $T_u=298$~K is considered. These conditions and fuel have been selected because both the hydrodynamic (Darrieus-Landau) and thermodiffusive instability are strong; in addition, these conditions have already been extensively studied~\citep{BERGER2022111935,BERGER2022111936}, making this a very well-understood case.
To obtain an accurate solution, a sufficient number of points $n_f$ should be distributed in the (thermal) flame thickness $\delta_f=(T_b-T_u)/(\partial T/\partial y)_{max}$, where $T_b$ and $T_u$ are the burnt and unburt temperatures, and $(\partial T/\partial y)_{max}$ is the maximum temperature gradient in a one-dimensional planar laminar flame. Therefore, a parametric analysis of the number of points is also included to understand the effect of resolution on the computed dispersion relation. Similarly to the resolution, the size of the computational domain can impact the dynamics of intrinsic instabilities~\citep{berger2019characteristic}, so this numerical parameter is also investigated. 
For the example presented in this section and used to explain the methodology, a domain with $L_x=384\delta_f$ is used. The mesh resolution is set to $n_f=25$ and the initial amplitude to $A_{ini}=2.3\cdot10^{-9}\delta_f$. As a comparison, analyses performed with the single-wavelength method usually use resolutions ranging from $n_f=10$ to $n_f=25$~\citep{YUAN20071267, altantzis_2012, FROUZAKIS20151087}. A summary of the parameters for the simulations shown in the paper is presented in Tab.~\ref{table:parameters}.

\begin{table}
    \centering
    {\footnotesize
    \begin{tabular}{lcccc}
    \hline
        Parameters & units  & value\\
    \hline
        fuel       & -      & H\textsubscript{2} \\
        $\phi$     & -      & 0.5     \\
        $T_u$      & K      & 298 \\
        $p$        & Pa     & 101325\\
    \hline
        $s_L$      & cm/s & 47.8  \\ 
        $\delta_f$ & m    & $4.39\cdot10^{-4}$  \\ 
        $\tau_f$   & s    & $9.99\cdot10^{-4}$  \\ 
    \hline
        $L_y$      & m    & $12\delta_f$\\
    \hline
        $n_f$      & -    & $[3,~25]$\\
        $L_x$        & m    & $[48,~384]\delta_f$\\
        $A_{ini}$  & m    & $[10^{-12},~10^{-3}]\delta_f$\\

    \hline
    \end{tabular}
    }
    \caption{Simulation and flame parameters: $\phi$ is the equivalence ratio of the gas mixture, $T_u$ is the unburnt gases' temperature, $p$ is the pressure, $s_L$ is the flame speed, $\delta_f$ is the flame thickness, $\tau$ is the flame time, $L_y$ is the domain length in the streamwise direction $y$, 
    $n_f$ is the number of point in the flame thickness, $L_x$ is the domain size (along the spanwise direction $x$), and $A_{ini}$ is the initial amplitude of the perturbation.}
    \label{table:parameters}
\end{table}

\subsection{Initial perturbation}

An initial perturbation, consisting of a sum of sinusoids of different wavelengths, is superimposed on the otherwise planar flame. For each spanwise location $x$, the 1D flame profiles are shifted in the streamwise direction $y$ with a linear interpolation by a distance:
\begin{equation}
    y_f(x, t_0)=\sum_{n=1}^{N} A_n(t_0)\sin\left(nx\frac{2\pi}{L_x}+\varphi_n\right)
\label{eq:pert}
\end{equation}
where $A_n(t_0)$ is the initial amplitude of the harmonics, $L_x$ the size of the domain, $N$ the desired number of harmonics $\lambda_n=L_x/n$, and $\varphi_n$ their phases. 
In all the simulations presented, the initial amplitudes of all the wavelengths are set to the same value $A_n(t_0) = A_{ini}$. While a more complicated initialisation, with different values of the initial amplitude for each wavelength, could be explored to obtain an optimal approach, we observed that the simple choice of a single value for all of them gives remarkably good results; in addition, the optimal choice is likely to be hard to guess a-priori, resulting in unnecessary complications. 
The 1D shift in Eq.~\ref{eq:pert} produces an initial condition that is not a solution to the governing equations. However, due to the very small values of $A_{ini}$ considered, this is not an issue since the initial unphysical transient is very short and well-behaved, as it will be shown in the following analysis.

\begin{figure}[h]
    \centering
    \includegraphics[width=\one]{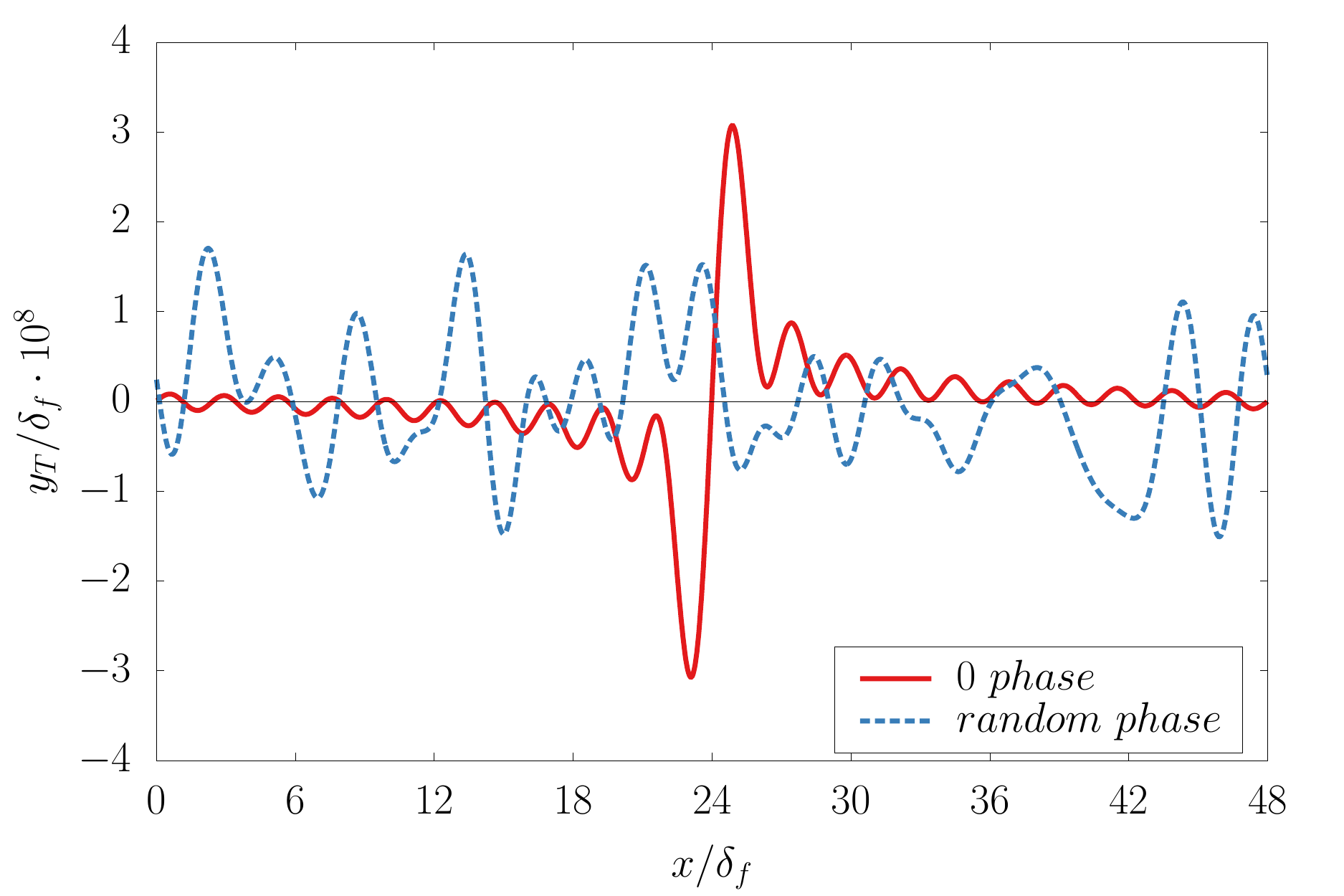}
    \caption{Initial harmonic perturbation for different values of the phases $\varphi_n$ in a domain of size $L_x = 48\delta_f$ and with initial amplitude $A_{ini}=2.3\cdot10^{-9}\delta_f$. The red solid line represents a perturbation with zero phases, $\varphi_n=0$, while the blue dotted line is generated with a random phase distribution. The lengths are normalised by the flame thickness $\delta_f$.}
    \label{fig:iso_random}
\end{figure}

Two examples of the initial condition of the perturbed flame are shown in Fig.~\ref{fig:iso_random} for two different sets of values of the phases $\varphi_n$. An initial profile obtained with random phases is compared with the profile computed with zero phases for all the wavelengths. In the rest of the analysis, initial perturbations with zero phases $\varphi_n=0$ are used without lack of generality, since the growth rate of the different wavelengths does not depend on the phase.
It is worth noting here that the values in the vertical axis of Fig.~\ref{fig:iso_random} are remarkably small, about eight orders of magnitude smaller than the flame thickness. As shown in the following, these small values guarantee a genuinely exponential growth of the instability, as expected in the linear regime.

The fundamental (minimum) and maximum wavenumber $k_1$ and $k_{max}$, as well as the maximum number of harmonics achievable $N_{max}$, are determined by the domain width $L_x$ and the mesh resolution. 
According to Shannon's theory~\citep{shannon1948mathematical}:
\begin{equation}
    k_{1} = \frac{2\pi}{L_x} = \frac{2\pi}{\lambda_1}
\end{equation}
\begin{equation}
    k_{max} = \frac{\pi}{\Delta x} = \frac{2\pi}{\lambda_{min}}
\end{equation}
\begin{equation}
    N_{max} = \frac{L_x}{2 \Delta x}
\end{equation}
where $\Delta x$ is the cell size in the spanwise direction $x$. The wavenumbers $k_n$ are related to the wavelengths as $k_n=2\pi/\lambda_n$. Therefore, for a given $\Delta x$, both $\lambda_{1}$ and $N_{max}$ increase with $L_x$. However, $k_{max}$ only depends on $\Delta x$.
Since the mesh size $\Delta x$ is small compared to the flame thickness $\delta_f$, the value of $k_{max}$ is always large enough to cover the entire range of interest for the dispersion relation. On the other hand, the size of the domain $L_x$ directly affects the minimum wavenumber $k_1$ and the resolution in wavenumber space of the dispersion relation. Therefore, it is important to assess the performance and robustness of the method for large domain sizes.

\subsection{Instability evolution and growth rates}

\begin{figure}[h]
    \centering
    \includegraphics[width=\one]{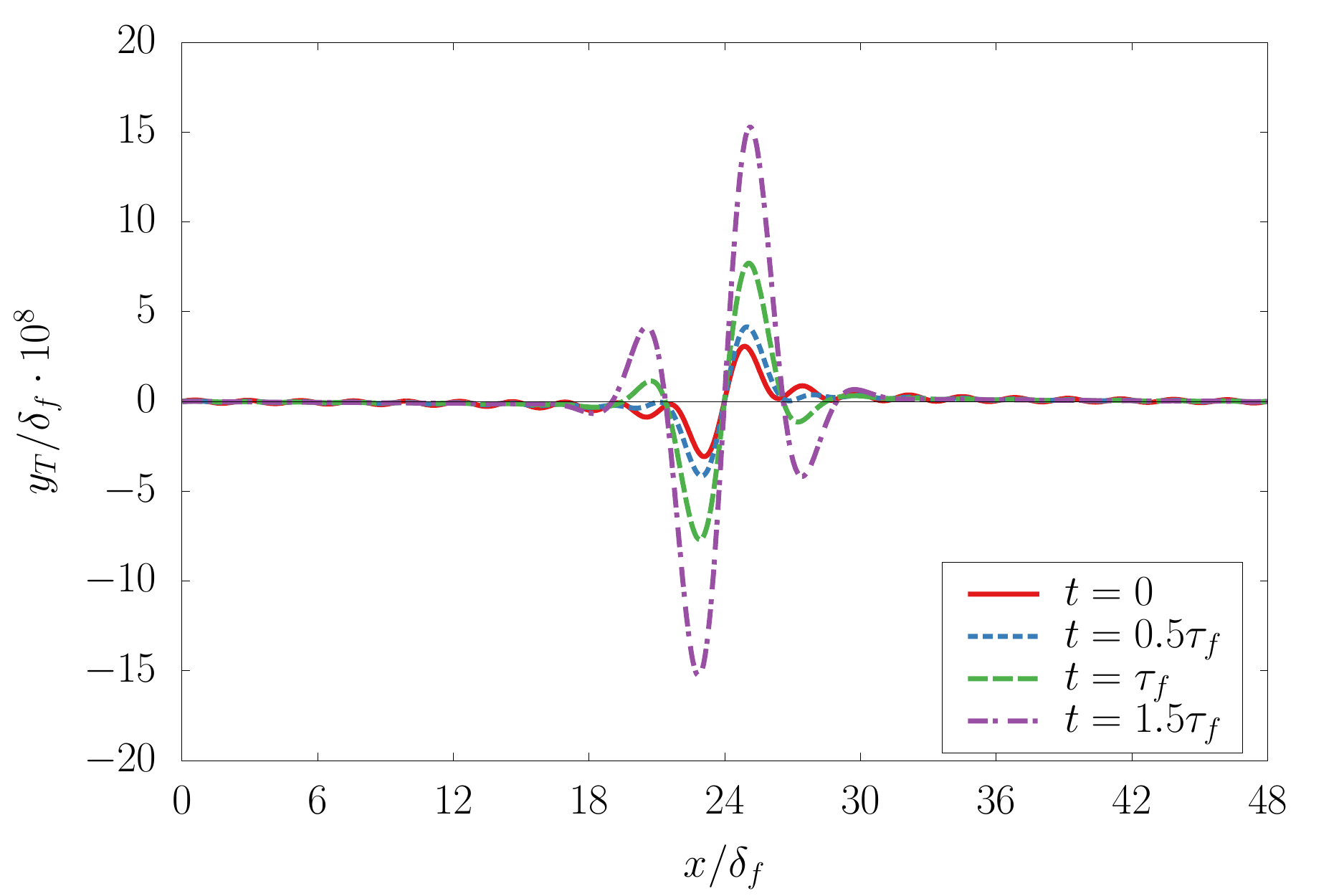}
    \caption{Evolution of the isoline of temperature $T=1000$~K over time for a planar flame perturbed with multiple wavelengths. In this case, $L_x = 48\delta_f$, $n_f=25$ and $A_{ini}=2.3\cdot10^{-9}\delta_f$. The time instant in the legend is normalised by the flame time $\tau_f=\delta_f/s_L$.}
    \label{fig:iso25}
\end{figure}

The amplitudes $A_n(t)$ of the perturbation components are monitored in time to capture the development of the flame instabilities. This is achieved by tracking the displacement of an isoline of temperature $y_T(x, t)$. In this paper, the analysis is based on the temperature field; other quantities, such as hydrogen mass fraction, have been considered without observing any difference. Fig.~\ref{fig:iso25} shows the isoline of $T=1000$~K at several time instants.  
The initial perturbation grows in a regular manner, with the most unstable wavelengths growing faster. Note that the peculiar shape shown in this figure is due to the choice of using zero phases in the initialisation; as mentioned before, this choice has no effect on the results. 

\begin{figure*}
    \centering
    \includegraphics[width=\two]{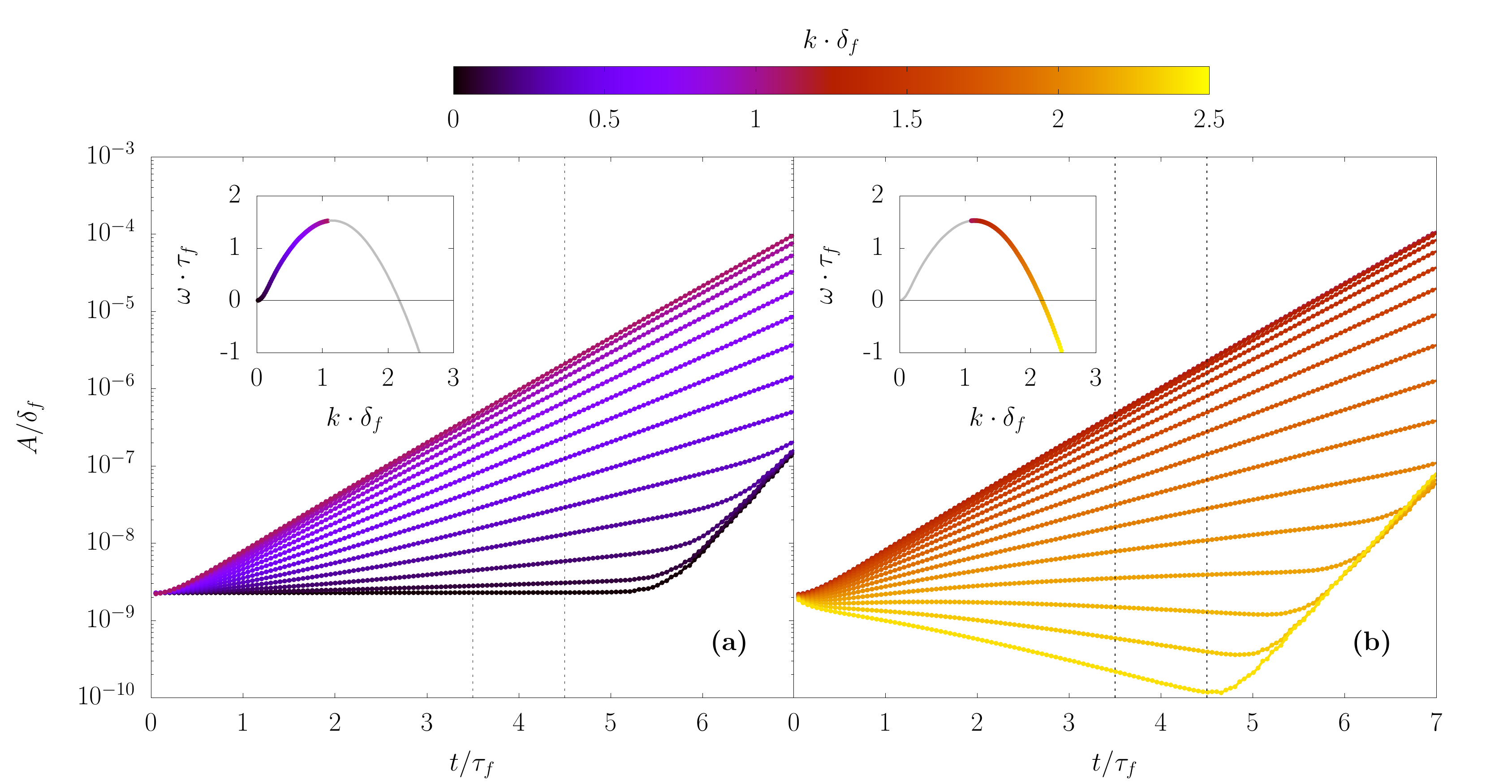}
    \includegraphics[width=\two]{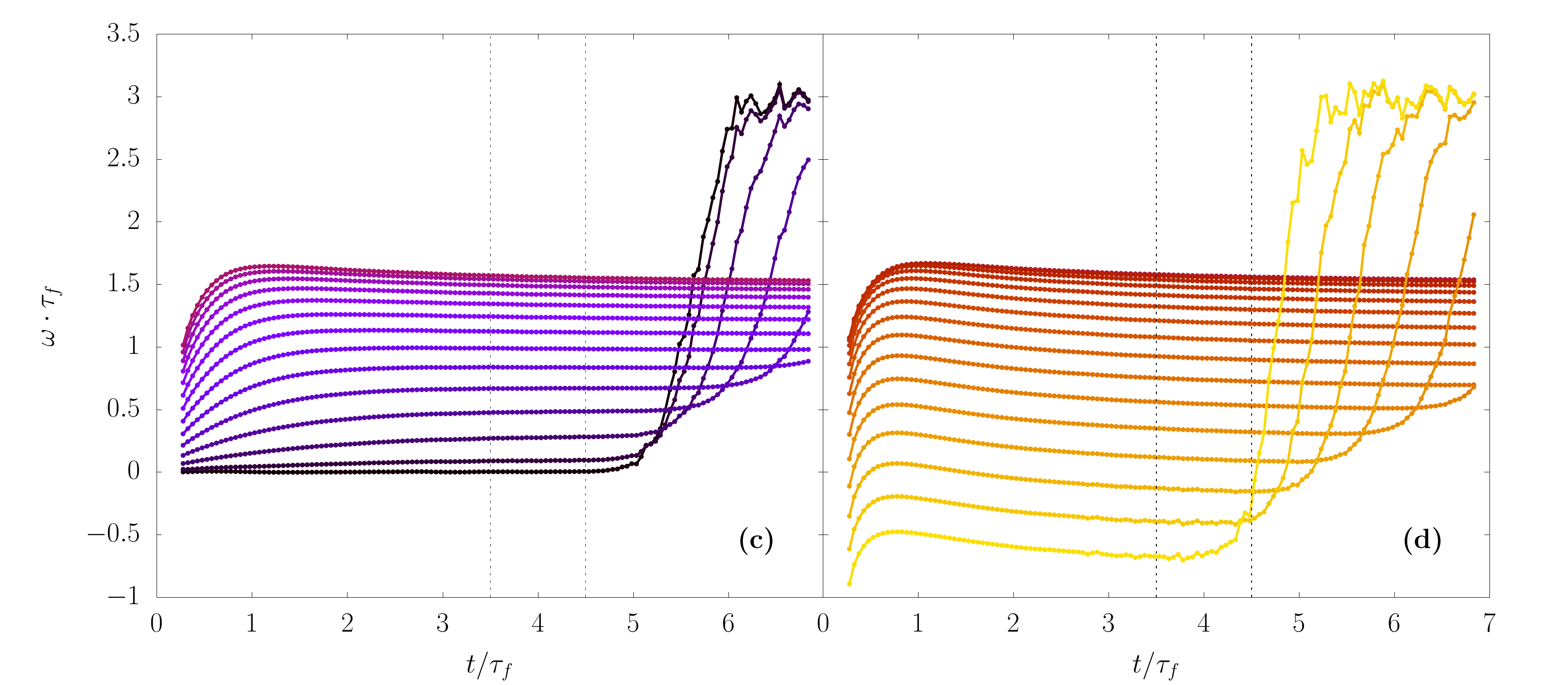}
    \caption{Evolution of the harmonics' amplitudes for $k<k_{\omega,max}$(a) and $k>k_{\omega,max}$ (b), and evolution of the harmonics' exponential growth rates for $k<k_{\omega,max}$(c) and $k>k_{\omega,max}$ (d). $L_x=384\delta_f$, $n_f=25$ and $A_{ini}=2.3\cdot10^{-9}\delta_f$. The vertical dotted lines delimit the averaging section used to compute $\omega_n$. The amplitude is normalised by the flame thickness $\delta_f$ and the time and growth rate by the flame time $\tau_f=\delta_f/s_L$, where $s_L$ is the laminar flame speed.}
    \label{fig:ampl160}
\end{figure*}

The amplitude of each wavelength is obtained by computing the discrete Fourier transform of $y_T$; Fig.~\ref{fig:ampl160}(a-b) shows the time evolution of the amplitude of 30 spectral components of the perturbation (out of the 160 computed). 
For clarity, the components with small and large wavenumbers are shown separately in two different panels.
The corresponding growth rates, shown in Fig.~\ref{fig:ampl160}(c-d), measure the rate of change of the amplitudes of the spectral components. The amplitudes follow an exponential growth $A_n(t)=A_n\exp(\omega_n t)$~\citep{BERGER2022111935}, where the exponential growth rate of the n-th harmonic can be determined as:

\begin{equation}
    \omega_n(t)=\frac{\mathrm{d}\ln\bigl(A_n(t)\bigr)}{\mathrm{d}t}
\end{equation}

After an initial transient, a linear phase is established. The logarithms of the amplitudes increase linearly, and each growth rate stabilises at a constant value. At later times, the values of $\omega_n(t)$ start to deviate as the flame evolution enters the non-linear phase, and the different wavelengths start to interact with each other. As shown in Fig.~\ref{fig:ampl160}, the harmonic components with large growth rates tend to stay longer in the linear regime compared to those with small and negative growth. In other words, intermediate $k$ stay longer in the linear regime than the larger and smaller ones. 
An interesting observation is that the harmonic components appear to deviate from the linear phase when the amplitude of other components, which initially had smaller growth rates but already transitioned to the non-linear regime, reach their amplitude. This suggests that the propagation of transition to non-linearity among the components happens sequentially, starting from the components with small growth rates and then progressively involving the components with larger growth rates, which are more resilient to transition. This may happen because low amplitude (and therefore low growth rate) components get affected by non-linear coupling with other higher amplitude (and thus higher growth rate) components.


\subsection{Dispersion relation}


The dispersion relation should be constructed by evaluating the growth rate of each component in a time range when the exponential growth rate is constant. 
Given the observations in the previous discussion, the time at which the growth rate is extracted should be selected with some care.
The appropriate time range can be selected after visually inspecting the evolution of the amplitudes and growth rates shown in Fig.~~\ref{fig:ampl160}. Another solution is to use a piecewise linear time series segmentation, or any other method that can detect the breaking points when the transition to non-linearity is observed in the evolution of the amplitudes. A unique averaging time range can also be chosen to simplify the analysis, as shown by the vertical lines in Fig.~\ref{fig:ampl160}. 

\begin{figure}[h]
    \centering
    \includegraphics[width=\one]{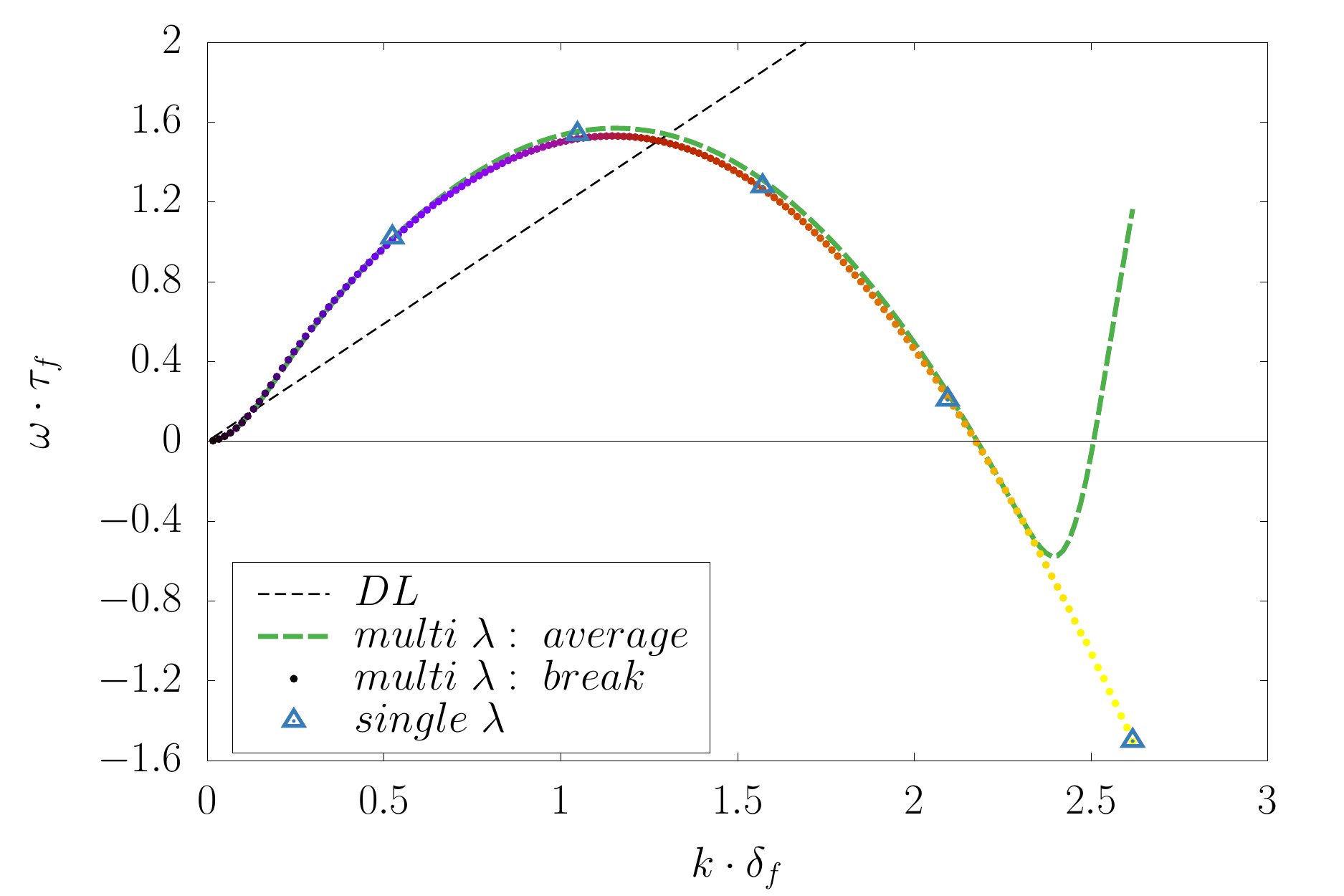}
    \caption{Dispersion relation for $L_x=384\delta_f$, $n_f=25$ and $A_{ini}=2.3\cdot10^{-9}\delta_f$. The thin black dashed line represents the growth rates associated with the Darrieus-Landau (DL) instability. The green dashed line and the multi-colour points represent the dispersion relation computed with the multi-wavelength approach points. They are respectively computed with a single averaging section and the breaking point detection method. The blue triangles represent the growth rates computed with the single-wavelength approach. The growth rate is normalised by the flame time $\tau_f$ and the wavenumber by the flame thickness $\delta_f$.}
    \label{fig:disp160}
\end{figure}

The growth rates are finally plotted as a function of the wavenumbers to represent the dispersion relation, as shown in Fig.~\ref{fig:disp160}. The results computed by averaging in the range marked by the vertical lines in Fig.~\ref{fig:ampl160} and the results obtained by identifying the breaking point give similar results for the lowest $k$. A minor deviation is observed in the range close to the maximum because the growth rates of the intermediate $k$ are not fully converged in the average range used. For very large $k$ (strongly negative growth rates) the results obtained with the average is different since these harmonics have already transitioned to the non-linear regime in the range selected. Anyway, the dispersion relation for values of $k$ larger than the cut-off is usually of little interest since these wavelengths are stable. Fig.~\ref{fig:disp160} also shows the theoretical growth rate for the Darrieus-Landau (DL) instability~\citep{matalon2007intrinsic}. This is included to highlight the typical behaviour in hydrogen flames; as expected, the dispersion relation is clearly above the DL line due to the presence of thermodiffusive instabilities. 

\section{Comparison between single- and multi-wavelength approaches}
\label{sec:comparison_single_multi}

A direct comparison between the new multi-wavelength approach and the traditional single-wavelength approach has been carried out to verify the accuracy of the proposed method. The results obtained with the single-wavelength approach are also shown in Fig.~\ref{fig:disp160}. For the single-wavelength approach, the growth rate of each wavelength is computed in a different simulation with a domain size matching the perturbation wavelength $L_x=\lambda_n$. The resolution is the same as that employed in the multi-wavelength approach. The amplitudes of the single-wavelength perturbation are determined by simply fitting the temperature isolines with sinusoids of wavelengths $\lambda_n$. More sophisticated methods based on the evaluation of the length of the isoline could also be employed, but it has been verified that the results are the same.
The similarity between the results obtained with the two approaches is clearly illustrated in Fig.~\ref{fig:disp160}. The growth rates computed with the single-wavelength approach match those computed with the multi-wavelength approach.

\begin{figure}[h]
    \centering
    \includegraphics[width=\one]{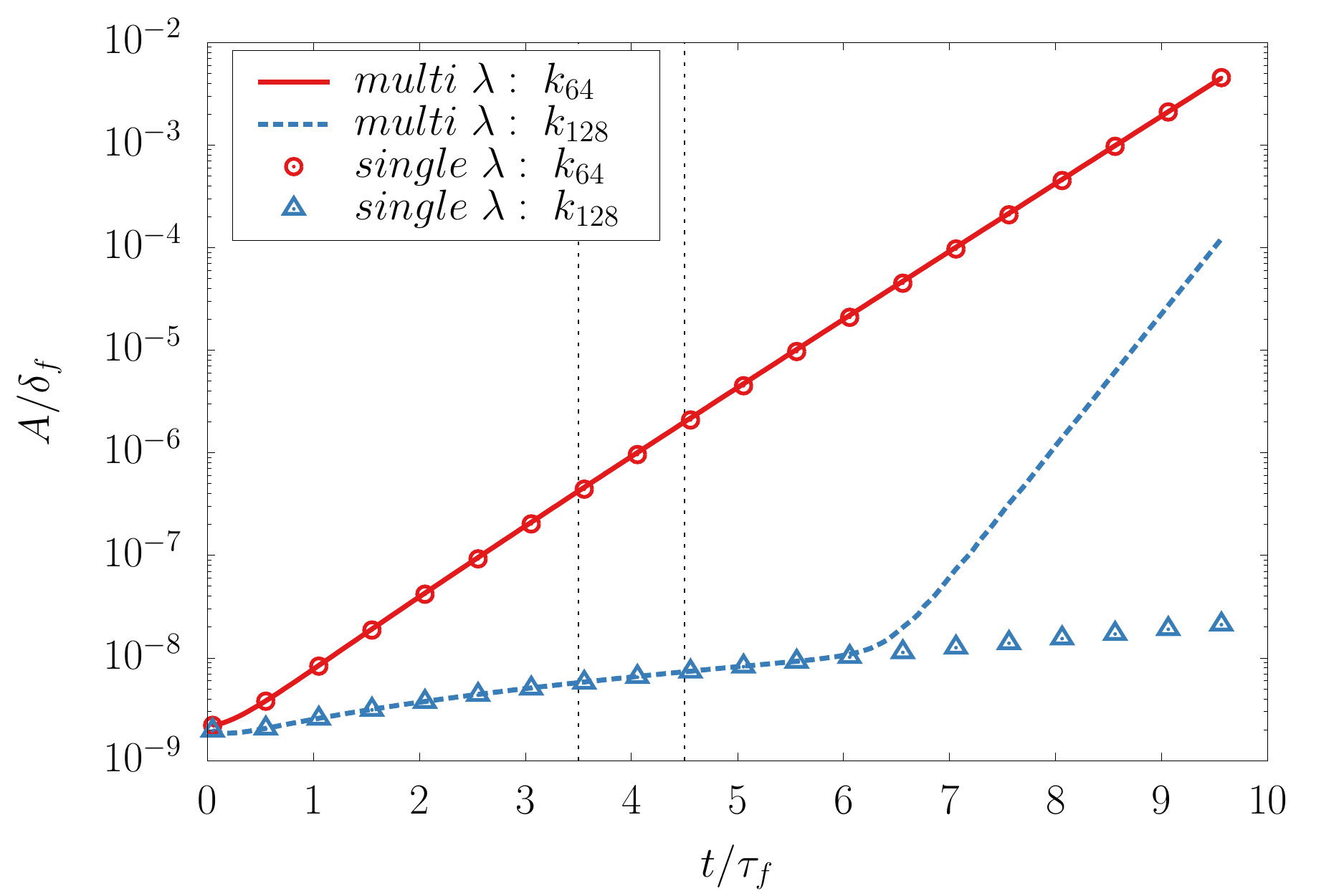}
    \caption{Comparison of the amplitude evolution for different wavelengths. $n_f=25$ and $A_{ini}=2.3\cdot10^{-9}\delta_f$. The lines and symbols are respectively obtained with the multi- and single-wavelength approach. The red and blue colours represent respectively the 64th harmonic $k_{64}\cdot\delta_f=1.05$ and 128th harmonic $k_{128}\cdot\delta_f=2.1$.}
    \label{ampl_ms}
\end{figure}

Fig.~\ref{ampl_ms} shows a comparison between the methods for the amplitude evolution of two selected perturbation components. The amplitude of the wavelength with a large growth rate ($k_{64}\cdot\delta_f=1.05$) behaves identically for the single- and multi-perturbation approaches. For the wavelength with a lower growth rate ($k_{128}\cdot\delta_f=2.1$), the interaction between the different harmonics triggers the transition to the non-linear regime earlier. However, their growth rates can be accurately computed, since a linear phase can be clearly identified, as shown in Fig.~\ref{fig:disp160}. 

\begin{figure}[h]
    \centering
    \includegraphics[width=\one]{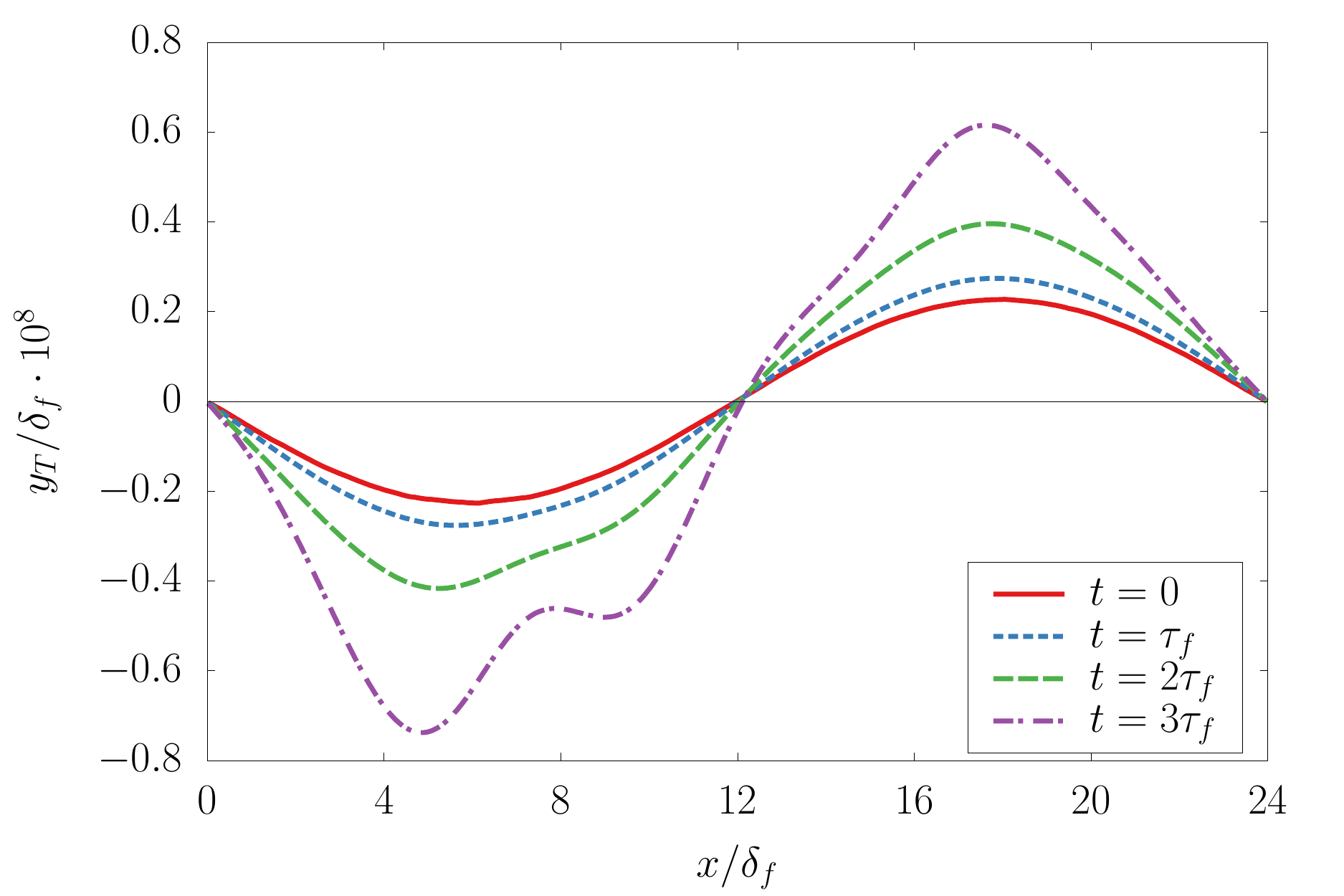}
    \caption{Evolution of an isoline of temperature $T=1000K$ over time for a single-wavelength perturbation with a small wavenumber $k\cdot\delta_f=0.26$ ($L_x = 24\delta_f$ $n_f=25$ and $A_{ini}=2.3\cdot10^{-9}\delta_f$).}
    \label{isok2}
\end{figure}

As mentioned in the introduction, an important aspect to consider is the difficulty often faced when dealing with small wavenumbers (much smaller than the wavenumber of peak growth) in the context of the single-wavelength approach. 
The evolution of the temperature isolines of a single large wavelength (small wavenumber) is shown in Fig.~\ref{isok2}. Due to the large domain size used in this case, perturbations with smaller wavelength (larger wavenumber) than that explicitly imposed as initial conditions might grow from the numerical noise. Since the growth rate of these is certainly larger than that of the single imposed one, they can easily become prevalent and compromise the assessment of the growth rate of the explicitly imposed wavelength. As shown in Fig.~\ref{fig:disp160}, this is not an issue for the proposed approach since the growth rates of the various wavelengths are disentangled by the spectral analysis, and the dispersion relation is properly computed also for very small $k$. This is important for hydrogen flames, characterised by thermodiffusive instabilities. These instabilities manifest themselves by a positive curvature of the dispersion relation near the origin, which is captured by the proposed method.

\section{Effects of numerical and initialisation parameters}
\label{sec:param}

\subsection{Resolution}

\begin{figure}[h]
    \centering
    \includegraphics[width=\one]{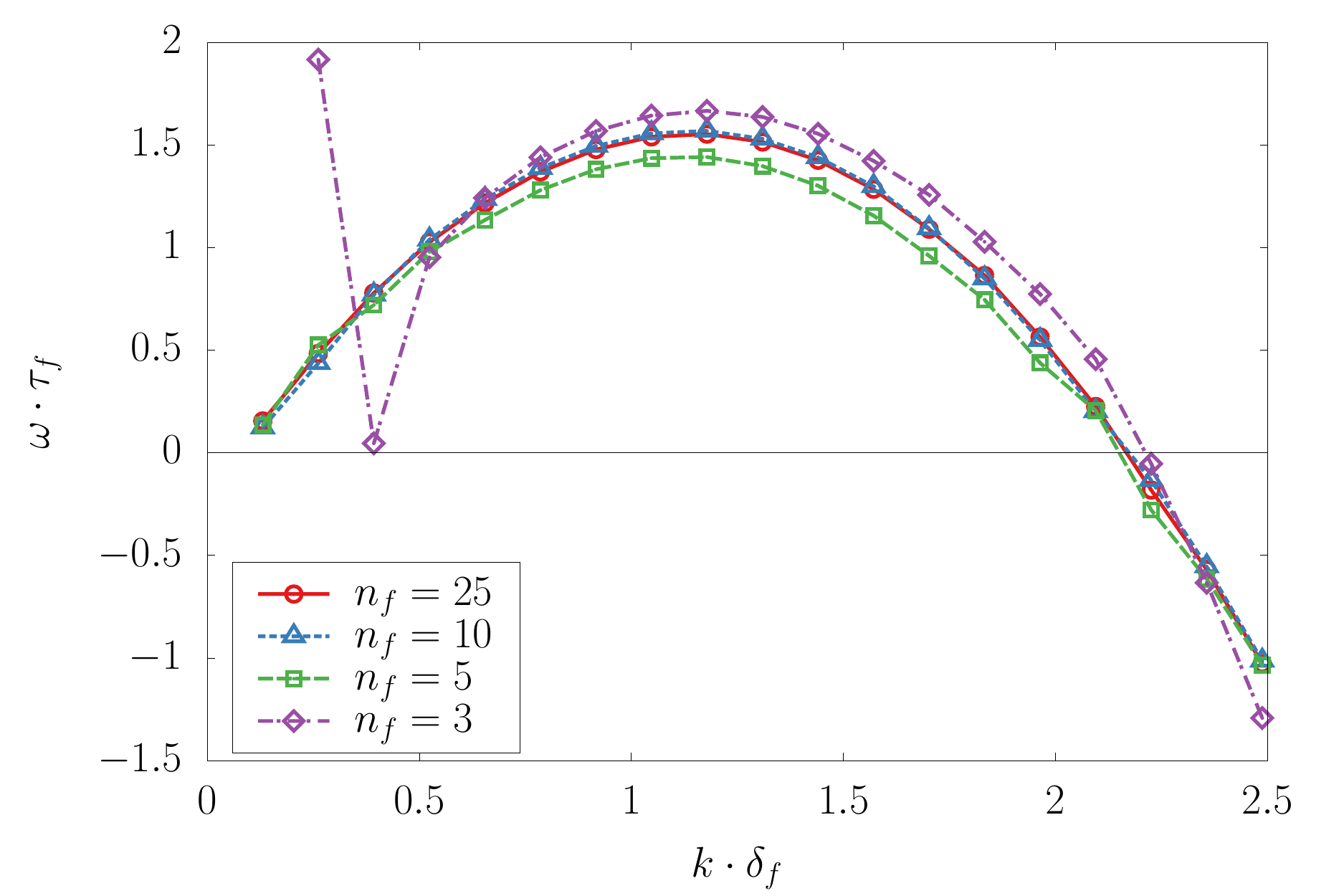}
    \caption{Dispersion relations for different mesh resolutions. $L_x=48\delta_f$ and $A_{ini}=2.3\cdot10^{-9}\delta_f$. The number of points in the flame thickness $n_f$ is related to the mesh resolution as $\Delta x=\delta_f/n_f$.}
    \label{fig:disp_nf}
\end{figure}

With the goal of ensuring an accurate computation of the dispersion relation and maintaining computational efficiency, it is essential to assess the 
resolution requirements needed with the proposed method. The number of points in the flame thickness $n_f$ is related to the mesh resolution as $n_f=\delta_f/\Delta x$.
We conducted a series of simulations with different mesh resolutions, ranging from $n_f=3$ to $n_f=25$.
The size of the domain is kept constant with $L_x=48\delta_f$, and the initial amplitude of the perturbation is set to $A_{ini}=2.3\cdot10^{-9}\delta_f$. The initial perturbation is generated with 20 wavelengths, from $\lambda_1=48\delta_f$ to $\lambda_{20} = 2.4\delta_f$. A comparison between the different mesh resolutions is presented in Fig.~\ref{fig:disp_nf}. The dispersion relation shows negligible differences between $n_f=25$ and $n_f=10$, which is consistent with the analyses performed with the single-wavelength method~\citep{YUAN20071267, altantzis_2012, FROUZAKIS20151087}. It deviates for smaller resolutions, so the accuracy for $n_f<5$ should be considered questionable.

\begin{figure}[h]
    \centering
    \includegraphics[width=\one]{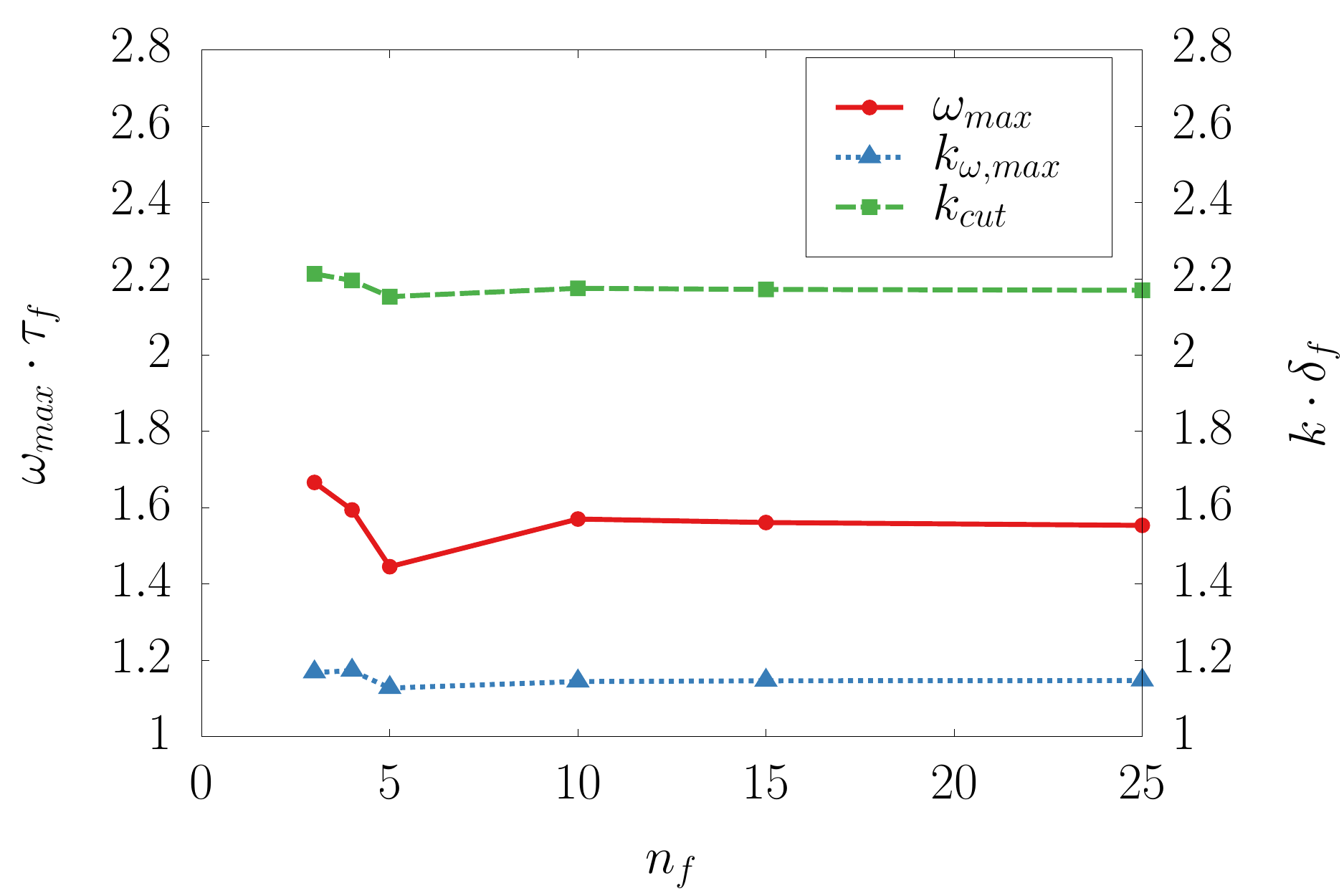}
    \caption{Effect of the mesh resolution on the maximum growth rate $\omega_{max}$, its wavenumber $k_{\omega,max}$  and the cutting wavenumber $k_{cut}$. $L_x=48\delta_f$ and $A_{ini}=2.3\cdot10^{-9}\delta_f$.}
    \label{fig:comp_nf}
\end{figure}

For modelling and theory development~\citep{matalon2007intrinsic}, a set of parameters characterising the dispersion relation is often of interest.
The wavelength of maximum growth $\lambda_{\omega,max}$ observed in the linear regime helps to characterise the size of the cellular structures in the non-linear regime~\citep{berger2019characteristic,BERGER2022111936}, while the maximum growth rate $\omega_{max}$ can be used to model the enhanced flame speed in the non-linear regime~\citep{HOWARTH2022111805,BERGER2022111936}.  
The cut-off wavelength $\lambda_{cut}$ is also an interesting feature~\citep{ATTILI20211973} that can help understanding if the instability could happen or not in a system of a given size~\citep{Lam2020}, and can also be of interest for modelling~\citep{LAPENNA20212001}; it corresponds to the wavelength that possesses a zero growth rate. Any wavelength smaller than $\lambda_{cut}$ or wavenumber larger than $k_{cut} = 2\pi/\lambda_{cut}$ will be dissipated out. 
These quantities are shown in Fig.~\ref{fig:comp_nf} for different resolutions.
The values of $\omega_{max}$, $k_{\omega,max}$ and $k_{cut}$ vary only slightly between $n_f=25$ and $n_f=10$. A trade-off between computation time and accuracy can be found around $n_f=10$ as the value of $\omega_{max}$ varies only by $1.1\%$, $k_{\omega,max}$ by $0.2\%$ and $k_{cut}$ by $0.2\%$ compared to the higher-resolution case. The resolution $n_f=5$ appears to be the lower limit to get a reasonable estimation. For smaller $n_f$, the dispersion relation is not computed accurately.

\subsection{Domain size}

\begin{figure}[h]
    \centering
    \includegraphics[width=\one]{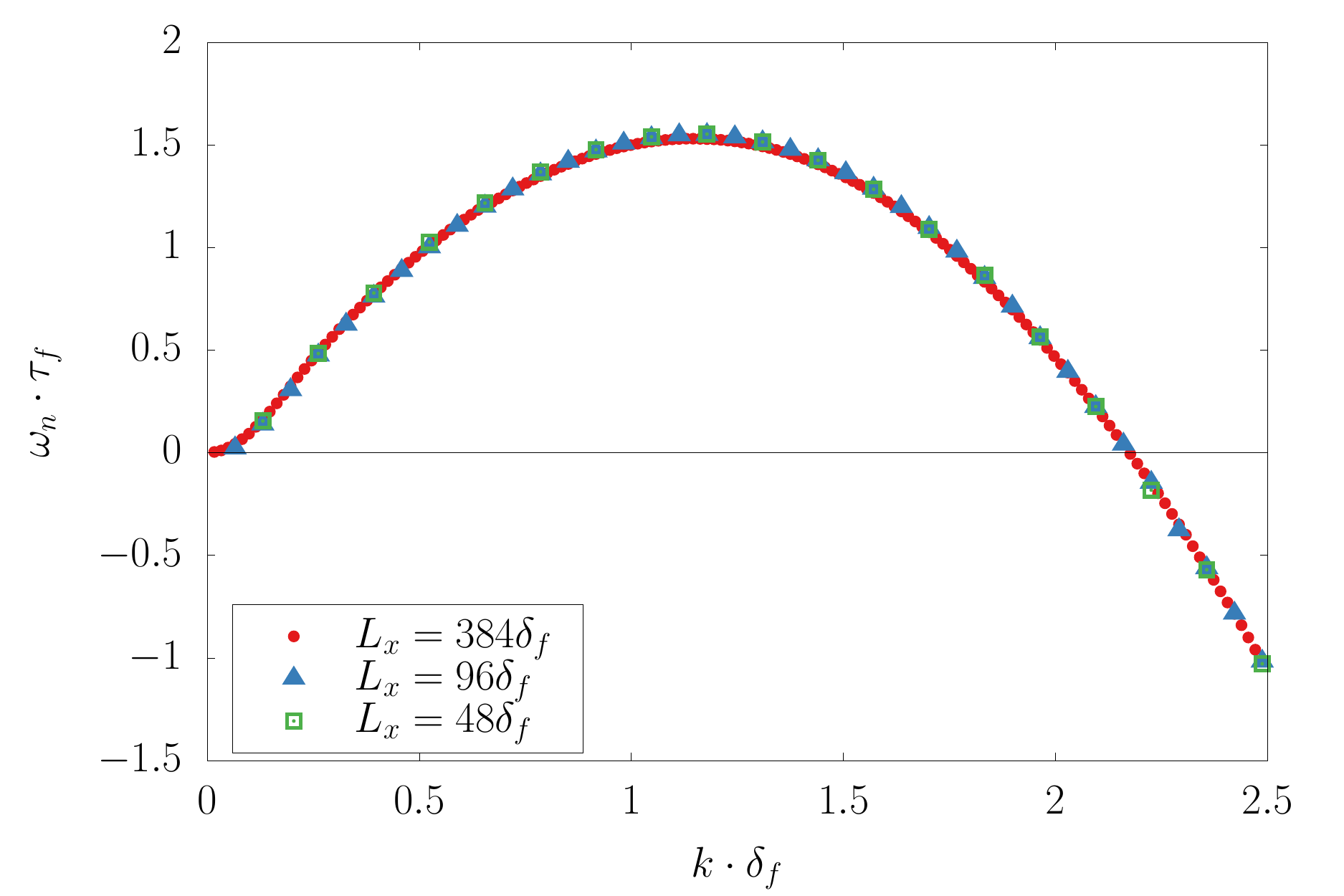}
    \caption{Dispersion relations for different domain sizes. $A_{ini}=2.3\cdot10^{-9}\delta_f$ and $n_f=25$.}
    \label{fig:disp_domain}
\end{figure}

The domain size is inversely proportional to the fundamental wavenumber $k_1=2\pi/L_x$, characterising the perturbation with the longest wavelength. With a larger domain, the dispersion relation is extended to smaller $k$ and the gap between two consecutive wavenumbers is reduced, increasing the number of points in the dispersion relation. A series of simulations with different domain sizes from $L_x=48\delta_f$ to $L_x=384\delta_f$ is performed to assess its effect. The initial amplitude of the perturbation is set to $A_{ini}=2.3\cdot10^{-9}\delta_f$, and the number of points in the flame thickness is $n_f=25$. The dispersion relations with the different domain sizes are compared in Fig~\ref{fig:disp_domain}. The results with different $L_x$ are negligibly different. As already mentioned, the resolution in wavenumber space increases and the dispersion relations extends to smaller $k$ if $L_x$ increases. These results show that the method is robust and the domain size can be chosen simply to get the desired resolution in $k$-space, without issues regarding accuracy and numerical stability. 

\subsection{Initial perturbation}

\begin{figure}[h]
    \centering
    \includegraphics[width=\one]{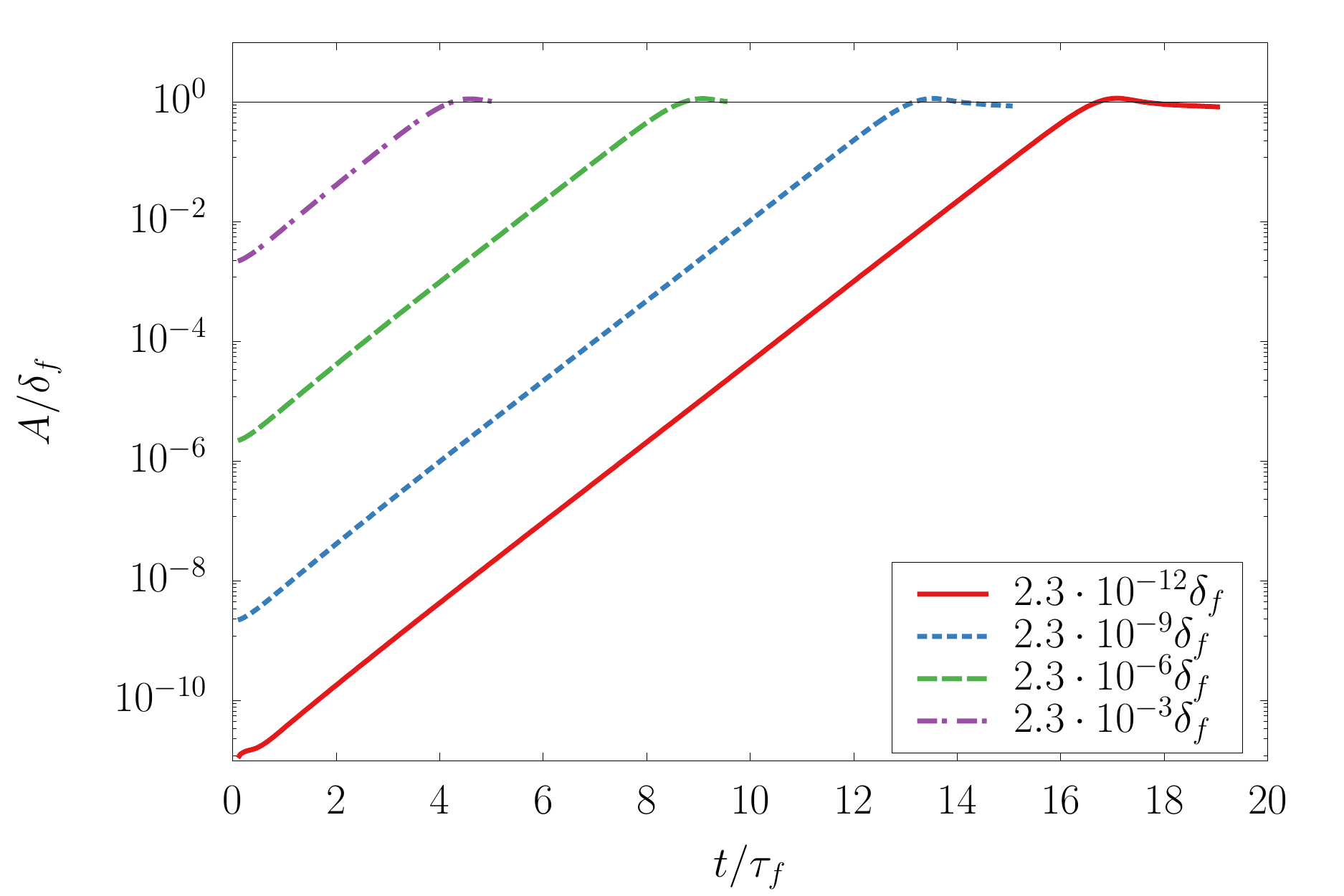}
    \caption{Amplitude evolution of single-wavelength perturbations with different initial amplitudes. $L_x=5.3\delta_f$ and $n_f=25$.}
    \label{fig:evolution_ampl}
\end{figure}

Another important parameter is the amplitude of the initial perturbation $A_{ini}$. It must be chosen wisely to ensure the existence of an initial linear regime that can be used to extract meaningful values of the growth rate. First, a series of simulations with different $A_{ini}$ from $2.3\cdot10^{-12}\delta_f$ to $2.3\cdot10^{-3}\delta_f$ are conducted with the single-wavelength approach. The domain size is set to $L_x=5.3\delta_f$, close to the maximum-growth rate wavelength. 
Fig.~\ref{fig:evolution_ampl} shows the time evolution of the amplitude of the single wavelength starting from a different initial amplitude. 
After a clean exponential growth, each case reaches a plateau and stabilises around the flame thickness $\delta_f$, independently of $A_{ini}$. The growth rate is the same for all cases. As expected, the duration of the linear phase is smaller when starting from a larger perturbation. 

\begin{figure*}
    \centering
    \includegraphics[width=\two]{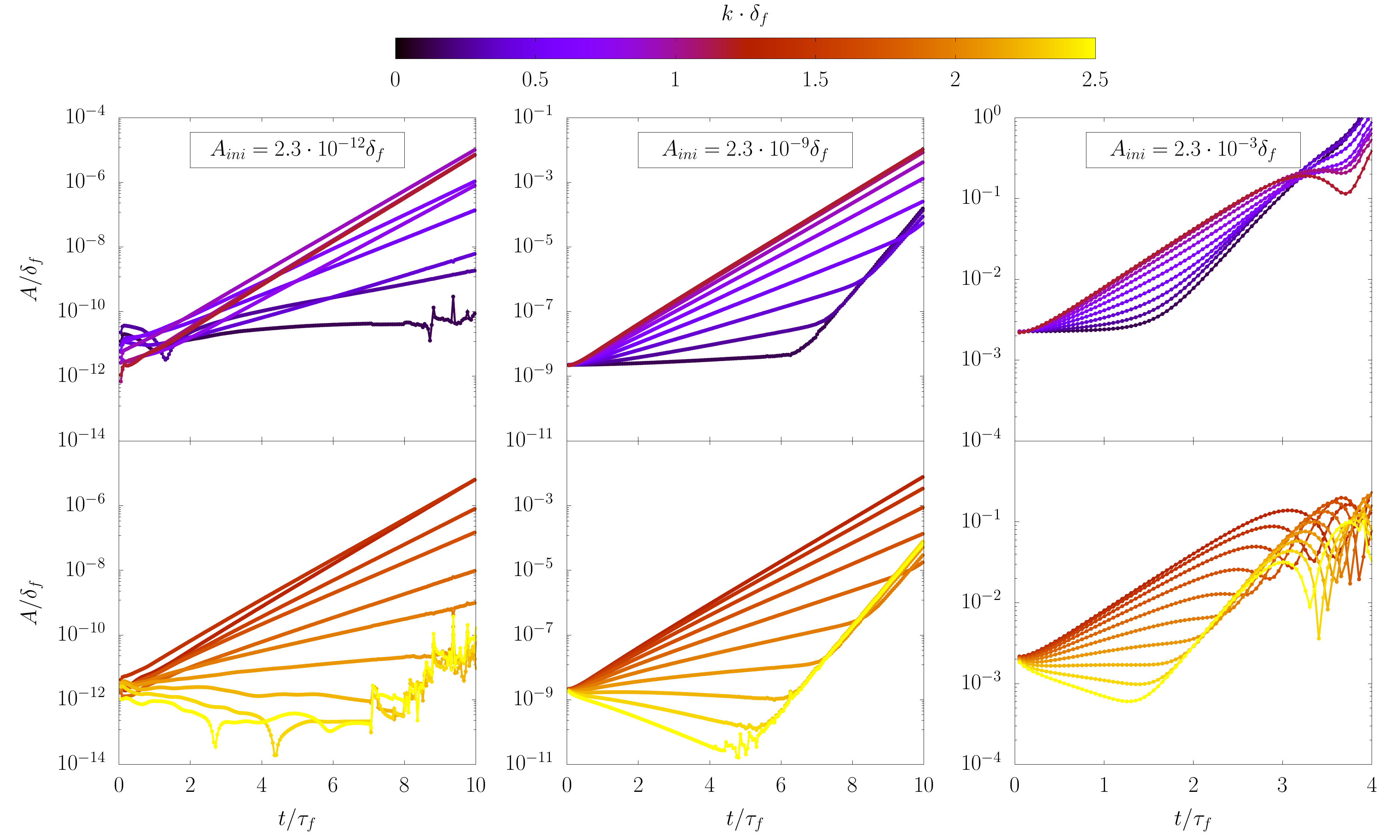}
    \caption{Evolution of the harmonic's amplitudes for different initial amplitudes. $L_x=48\delta_f$ and $n_f=25$. Small (large) wavenumbers are shown in the top (bottom) row.}
    \label{fig:ampl_aini}
\end{figure*}

The analysis is then extended to the case of the multi-wavelength approach, performing a series of simulations with different initial amplitudes ranging from $A_{ini}=2.3\cdot10^{-12}\delta_f$ to $A_{ini}=2.3\cdot10^{-3}\delta_f$. The domain size is set to $L_x=48\delta_f$ and the grid resolution to $n_f=25$. The initial perturbation is generated with 20 wavelengths from $\lambda_1=48\delta_f$ to $\lambda_{20} = 2.4\delta_f$. 
Fig.~\ref{fig:ampl_aini} shows the time evolution of the various wavelengths for three different values of the initial amplitude.  
For a large value of the initial perturbation, it is difficult to identify a time range long enough to compute accurate growth rates. In particular, the linear regime of the low-amplitude harmonics is too short to compute their growth rates accurately. On the other hand, $A_{ini}$ should not be too small to initialise the perturbation correctly. The initial transient before the linear growth is achieved is long, and the initial amplitude is no longer imposed by the initial simulation settings but by the numerical noise, making the evolution less clear. 

\begin{figure}[h]
    \centering
    \includegraphics[width=1\one]{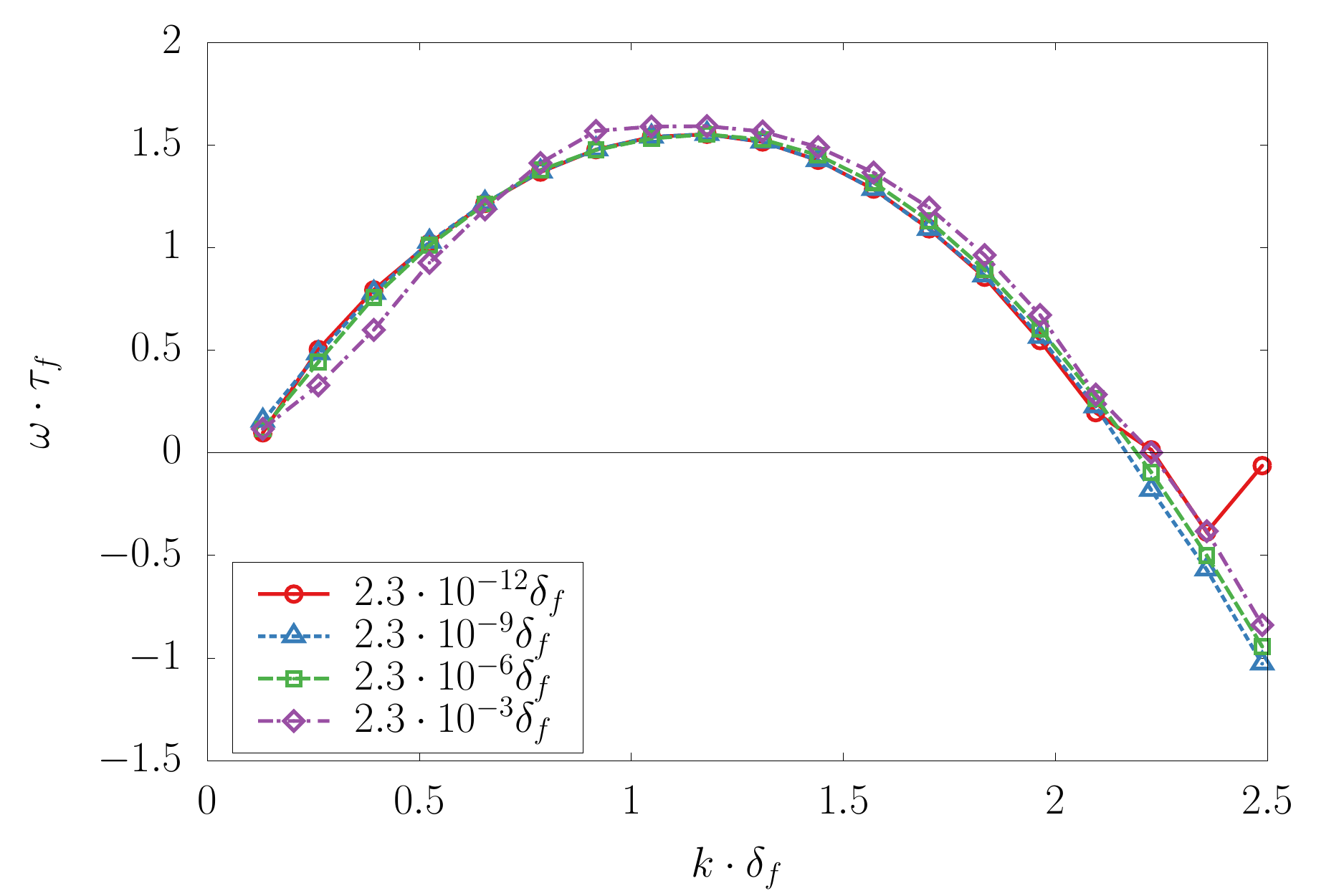}
    \caption{Dispersion relations for different initial amplitudes. $L_x=48\delta_f$ and $n_f=25$.}
    \label{fig:disp_pert}
\end{figure}

A comparison between the dispersion relation obtained with the different initial perturbations is presented in Fig.~\ref{fig:disp_pert}. The dispersion relations do not vary between $A_{ini} = 10^{-9}\delta_f$ and $10^{-6}\delta_f$. The dispersion relation for $A_{ini} = 2.3\cdot10^{-3}\delta_f$ appears to be less accurate due to the difficulty of isolating a proper time interval characterised by linear dynamics. For very small perturbations, an issue is observed for large wavenumbers; in this case, the very long transient required to establish the linear regime limits the extent of the linear phase. 
A proper balance between the initialisation of the perturbation and the length of the linear regime is found in the range $A_{ini} \in [10^{-11},~10^{-8}]\delta_f$.

\subsection{Interpolation of temperature isolines}

\begin{figure}[h]
    \centering
    \includegraphics[width=\one]{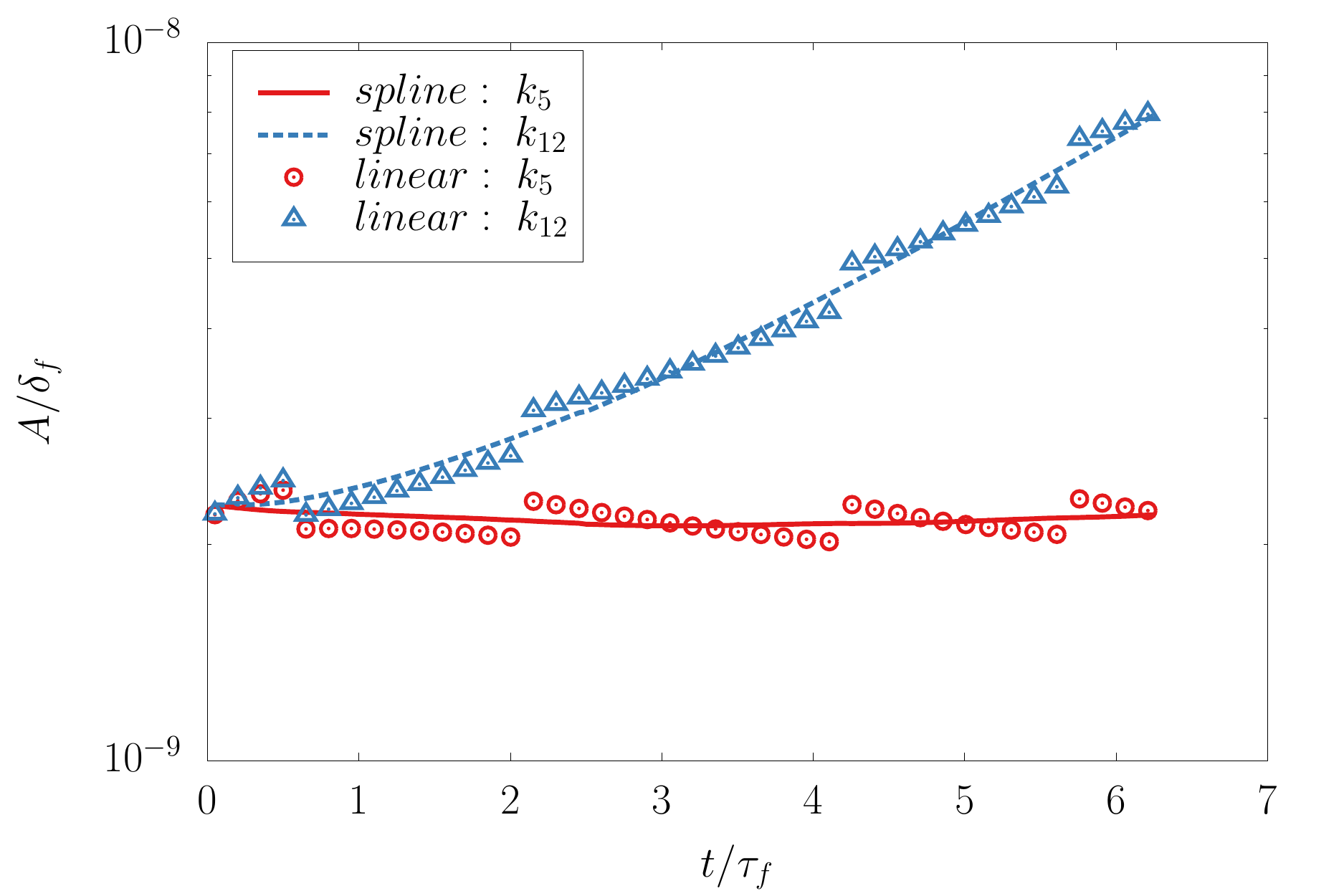}
    \caption{Comparison of the amplitude evolution for different interpolation methods to extract the isoline. $L_x=384\delta_f$, $n_f=10$ and $A_{ini}=2.3\cdot10^{-9}\delta_f$. The lines and symbols are respectively obtained with the spline and linear interpolation. The red and blue colour represents the 5th harmonic $k_{5}\cdot\delta_f=0.082$ and 12th harmonic $k_{12}\cdot\delta_f=0.196$, respectively.}
    \label{fig:interpolation}
\end{figure}

The identification of the isolines of temperature is crucial to properly compute the growth rate. Typical approaches to extract contours, such as the \emph{contourc} function in Matlab~\citep{MATLAB}, employ some type of interpolation to define the location of the points of the domain where the field has the specified value. The interpolation is often performed with a linear approximation. It was found that the use of a linear interpolation to extract the contour could lead to inaccuracies in the computation of the perturbation amplitude, and that a more sophisticated method should be employed. Fig.~\ref{fig:interpolation} shows the time evolution of the amplitude computed from temperature isolines extracted with linear and spline interpolation.  
It is evident that the more accurate interpolation approach is needed to avoid spurious jumps that would compromise the calculation of the growth rate.


\section{Method efficiency}
\label{sec:eff}
As explained in the introduction, this method aims to reduce the time required to compute a complete dispersion relation. Given a maximum wavelength $L_x$ and a number of points in the dispersion relation $N$, $N$ simulations of domain sizes $L_x/n$ where $n \in [1, N]$ have to be performed to compute the whole dispersion relation with the single-wavelength perturbation method. In contrast, only one simulation of domain size $L_x$ is required for the multi-wavelength approach. Therefore, the number of CPU hours needed for the single-wavelength approach evolves as the harmonic series, where the N-th harmonic number $H_N$ is given by:
\begin{equation}
    H_N = \frac{1}{1}+\frac{1}{2}+\frac{1}{3}+\cdots+\frac{1}{N} = \sum_{n=1}^{N} \frac{1}{n}
\end{equation}
It can be approximated by~\citep{harmonic}:
\begin{equation}
    H_N = \ln{N} + \gamma + o(1)
\end{equation}
where $\gamma \approx 0.58$ is the Euler's constant. Therefore, the number of CPU hours to compute a 20 points dispersion relation with the multi-wavelength perturbation method is a factor of $3.5$ smaller in comparison with single-wavelength approach. However, it should be borne in mind that this factor does not include the time required to prepare all the simulations, nor the possible problems that might be encountered when trying to find a specific point, such as the maximum growth rate of the cut-off wavelength. The real benefit of this method is the automation of the  dispersion relation computation.

\section{Conclusion}
\label{sec:concl}
An approach based on the spectral analysis of the time evolution of isolines of perturbed two-dimensional flames
is proposed to compute the whole dispersion relation of unstable flames with a single simulation. The method is computationally efficient since it avoids the need to perform multiple simulations, such as when the dispersion relation is computed point-by-point with a simulation for each wavelength. The effects of several numerical and initialisation parameters have been analysed. It is shown that a resolution of about ten grid points per flame thickness is enough and appropriate to obtain accurate results. The size of the domain has also been assessed; the method is robust, so the size of the domain can be selected based on the desired resolution of the dispersion relation in wavenumber space, without observing issues in the accuracy depending on the domain size. 
The importance of the initial amplitude of the imposed perturbation has been investigated, and it is shown that very large and very small values should be avoided. Still, a large range of initial perturbation amplitudes, of more than six orders of magnitude, provides accurate results. Finally, it is emphasised that the isolines should be computed with high-order interpolation methods to guarantee a precise computation of the growth rates.

\section{Acknowledgments}

P.E.L. acknowledges the support of Sapienza University by means of the early-stage researchers' funding.
H.P. and L.B. acknowledge funding by the European Union (ERC, HYDROGENATE, 101054894). Views and opinions expressed are however those of the author(s) only and do not necessarily reflect those of the European Union or the European Research Council Executive Agency (ERCEA). Neither the European Union nor the granting authority can be held responsible for them.



\bibliographystyle{elsarticle-num.bst}
\bibliography{biblio} 


\end{document}